\documentclass[preprint,aps]{revtex4}
\usepackage{graphicx}
\begin{document}
\title{Active-Sterile Neutrino Conversion: Consequences for the
$r$-Process and Supernova Neutrino Detection}
\author{J. Fetter,$^{1}$ G. C. McLaughlin$^{2}$,  
A. B. Balantekin,$^{1}$  and G. M. Fuller,$^{3}$}
\affiliation{
$^{1}$Department of Physics, University of Wisconsin, Madison,
Wisconsin 53706\\
$^{2}$Department of Physics, North Carolina State University, Raleigh,
NC 27695-8202\\
$^{3}$Department of Physics, University of California, San Diego,
La Jolla, CA, 92093-0319\\
}
\date{\today}

\begin{abstract}
We examine active-sterile neutrino conversion in the late time
post-core-bounce supernova environment.  By including the effect of
feedback on the Mikheyev-Smirnov-Wolfenstein (MSW) conversion
potential, we obtain a large range of neutrino mixing parameters which
produce a favorable environment for the r-process.  We look at the
signature of this effect in the current generation of neutrino
detectors now coming on line. We also investigate the impact of the
neutrino-neutrino forward scattering-induced potential on the MSW
conversion.

\end{abstract}
\maketitle

\newpage
\section{Introduction}
\label{sec:intro}

One form of element synthesis which may take place in core collapse
supernovae ({\it e.g.,} Type II, Ib, Ic) is the $r$-process, or rapid
neutron capture process.  The
$r$-process of nucleosynthesis has long been thought to be the
mechanism for producing many of the heavy nuclei (mass number, A $>$
100) \cite{bbfh}.  Successful production of the
r-process elements with anything approaching a solar system-like
abundance distribution requires a
neutron-rich environment, at least in the conditions
suggested by current supernova models and simple neutrino-driven wind
models without extremely rapid outflow and/or extremely relativistic
neutron star configurations.

In this paper, we study a primary $r$-process which occurs in stages.
The material in a fluid element moving away from the neutron star
remnant left by a core collapse supernova explosion
will experience three stages of nuclear evolution.
First, only free nucleons are present, then alpha particles coexist with
free neutrons.  At still lower temperatures the matter is
composed of alpha particles,
  ``seed'' nuclei with $50 \lesssim A \lesssim 100$, and
neutrons.  In the last stage the neutrons capture on the seed nuclei.

The neutrino-driven-wind environment, which is thought to occur at late
time (time post bounce $t_{pb} \sim 10 {\rm s}$) in the supernova is a
promising candidate site for the production of the r-process elements
\cite{r1}.  The wind is thought to arise well after the initial
collapse/explosion event; its successful generation of course
presupposes a successful explosion.  In the classic neutrino driven wind
models, neutrinos transfer energy to material near the surface of the
protoneutron star.  As this material flows outward, it expands and
cools.  As it cools, weak and electromagnetic/strong nuclear reaction
rates drop out of equilibrium sequentially.  If enough neutrons are
present after/during this freeze-out process, the $r$-process may take
place.  The number of neutrons is determined by three factors, the
initial neutron-to-proton ratio, the outflow timescale and the entropy of
the material.  Semianalytic models of the neutrino driven wind have shown
that it is difficult, though perhaps not impossible, to generate the
required numbers of neutrons per seed nucleus \cite{hwq}.

Self-consistent inclusion of neutrino capture reactions into a reaction
network code has shown that the problem of obtaining the required number of
neutrons is exacerbated by the interplay between neutrino and nuclear
reactions, especially the formation of alpha particles as described below
\cite{mmf}.  This is the so-called \lq\lq alpha
effect.\rq\rq\ The relative number of neutrons and protons is usually cast
in terms of the electron fraction,
$Y_e = 1/(1 + n/p)$, the net number of electrons (electrons minus
positrons) per baryon.
The electron fraction
 is determined mostly by the rates of neutrino and antineutrino
capture on free neutrons and protons \cite{Qetal}.  The most detrimental
effect, the alpha effect, takes place when material which would
eventually undergo heavy element nucleosynthesis passes through the
intermediate step of forming alpha particles.  At this stage, all the
protons are locked into alphas, but any excess neutrons remain free.
Neutrino captures on the remaining neutrons decrease the ratio of
neutrons to protons (or equivalently raise the electron fraction).
Supernova neutrino energies are too low to permit compensating charged
current captures on alpha particles \cite{mmf}.
Without sufficient numbers of neutrons, there can be no rapid neutron
capture process in these sorts of wind scenarios.

There are three potential solutions to this problem.  One solution involves
hydrodynamics.  For example, a very fast outflow may in principle cure the
problems associated with this environment \cite{hwq} and also the alpha
effect \cite{cf}.
Alternately, other scenarios which have rapid expansion (perhaps
followed by a slow down) accompanied by  high entropy and temperature 
may also cure the problem \cite{r1}.  However, it is not clear that 
such a fast outflow
or high entropy is present in this environment, or that either can be
achieved with simple neutrino heating.
 
The second solution may be to look for another site, such as neutron star
mergers.  Some neutron rich material may be ejected in the merger and
this has been shown to be capable of
producing r-process elements \cite{bsm89}.  These
mergers are likely too rare to reproduce the total observed abundance
of r-process elements, unless a significant fraction of the neutron
star material is ejected \cite{thielemann}. There is significant
energy released in the form of neutrinos in these models.  
Therefore, the alpha effect may 
compromise the r-process in these environments as well, although the
neutrino average energies in neutron star -neutron star mergers could
be  somewhat lower than in core collapse supernovae and neutron star -
black hole mergers, as discussed in e. g. Ref. \cite{janka99}.

A third solution is the one that is investigated in this paper:
active-sterile ($\nu_e \leftrightarrow \nu_s$, $\bar{\nu_e}
\leftrightarrow \bar{\nu_s}$) neutrino transformation.  The $\nu_s$ in
our study is defined as a particle which mixes with the $\nu_e$ (and
possibly also with $\nu_\mu$, and/or $\nu_\tau$) but does not contribute to
the width of the Z boson.  One candidate for this particle, although
not the only one, is a right handed Dirac neutrino.  A large number of
models exist for such light ``sterile'' neutrinos, for example, any
SU(2) Standard Model singlet.

In fact, the existence of sterile neutrinos with large masses ({\it
e.g.,} of order the unification scale) would not be unexpected, as they
are suggested by many neutrino mass/mixing models. However, light
sterile neutrinos do not often occur naturally in these models. 

The Sudbury Neutrino Observatory
(SNO) and the SuperKamiokande (SuperK) experiments together observe high
energy solar neutrinos and SuperK can observe and characterize the
atmospheric neutrino flux. What is emerging is a picture in which 
there is near
maximal mixing between the mu and tau flavor neutrinos at the atmospheric
neutrino mass-squared difference scale $\delta m^2 \approx
3\times{10}^{-3}\,{\rm eV}^2$, and a convincing result that two-thirds of
the neutrinos coming from the sun are mu/tau neutrinos, narrowly favoring
again near maximal mixing between the electron flavor neutrinos produced
by nuclear reactions in the sun and mu/tau flavor neutrinos at the solar
neutrino mass-squared difference $\delta m^2 <{10}^{-4}\,{\rm eV}^2$
\cite{superk,sno,bah}.  Is there any indication of light sterile degrees of 
freedom?

The Los Alamos Liquid Scintillator Neutrino Detector (LSND) experiment
remains the only possible indication of neutrino flavor oscillations (in
the $\nu_{\mu} \rightleftharpoons\nu_e$ channel at a neutrino
mass-squared scale $\delta m^2 \sim 0.5\,{\rm eV}^2$) 
beyond those derived from solar and atmospheric
neutrino considerations \cite{lsnd}. 
The mini-BooNE experiment at Fermilab will soon
check this result. If LSND holds up then it is tantamount to direct
evidence for the existence of a light sterile neutrino. Four neutrino
schemes designed to fit all of the existing neutrino oscillation data 
\cite{gg} are, however, barely viable or are outright challenged by the data. 
By contrast, we emphasize here that our considerations of the effect of a
sterile neutrino on supernova physics are completely independent of the
LSND result. For example, the sterile neutrino mass range which we find to
be efficacious in supernova r-process nucleosynthesis spans the entire
LSND-inspired range but also extends to considerably higher values of
$\delta m^2$ and to smaller vacuum mixing angles.

The effect of active-sterile transformation differs from that of
active-active ({\it e.g.}, $\nu_e \leftrightarrow \nu_{\mu\tau}$)
transformation, in that active-active transformation usually tends to
decrease the number of neutrons in the supernova, while active-sterile
transformation can increase it. (Active-active transformation can
slightly decrease the electron fraction for certain narrow ranges of
neutrino mass-squared difference.)  It has been shown that matter-enhanced
$\nu_e
\leftrightarrow \nu_{\mu,
\tau}$ could under some circumstances rather drastically affect the
neutron to proton ratio in the neutrino-driven wind and, thereby, affect
the prospects for the r-process there
\cite{Qetal}.

Active-sterile transformation in the supernova environment must be
treated carefully for several reasons.  The first is that, due to the
structure of the Mikheyev-Smirnov-Wolfenstein (MSW) conversion
potential, both neutrinos and antineutrinos may undergo transformation
in the same scenario.  In fact, in the solutions discussed here  both
types of transformations take place.  The second is that the neutrinos
themselves in part determine the MSW potential.  Any calculation of
the effect of oscillations must therefore include a feedback loop.  The
neutrinos determine the electron fraction; the electron fraction
determines the potential; and the potential determines the spectrum of
neutrinos.  The spectrum of neutrinos then sets the electron fraction.
Additionally, the $\nu \nu$-forward scattering contribution to the
MSW potential depends on the flavor states of the neutrinos.

In section \ref{sec:mech} we review the calculations which illustrate the
feedback mechanism \cite{us}.  Our study differs from a previous
study \cite{nunokawa} in that we track the thermodynamic and nuclear
statistical equilibrium evolution (NSE) of a mass element and update the
numbers of neutrons and protons at each time step directly from the weak
capture rates. This is essential to
accurately determine the number of neutrons available for the r-process.
In addition to producing a favorable $n/p$ ratio, this type of
transformation can suppress the population of $\nu_e$ and thereby defeat
the alpha effect.  In section
\ref{sec:detect} we discuss the observational consequences of such a
neutrino flavor transformation scenario in current neutrino detectors.
In section
\ref{sec:back} we explore the importance of neutrino background terms, or
neutrino-neutrino scattering in the MSW potential.  In section
\ref{sec:concl} we give conclusions.

\section{The Mechanism}
\label{sec:mech}

In this section, we  briefly describe our calculations and then we
summarize the impact of active-sterile neutrino transformation on the
electron fraction in the neutrino driven wind.  Our study in this
paper differs from that in our earlier work in Ref.\ \cite{us} in that
here we discuss the expected effects of active-sterile neutrino flavor
transformation on the expected supernova neutrino signal in several
different detectors and we also consider effects of the neutrino-neutrino
forward scattering-induced potential (the \lq\lq neutrino
background\rq\rq) on the neutrino flavor transformation problem.

If we neglect $\nu$-$\nu$ forward scattering contributions to the
weak potentials, then the equation which governs the evolution of the
neutrinos as they pass though the material in the wind can be written as:

\begin{equation}
i\hbar \frac{\partial}{\partial r} \left[\begin{array}{cc} \Psi_e(r)
\\ \\ \Psi_s(r) \end{array}\right] = \left[\begin{array}{cc}
\varphi_e(r) & \sqrt{\Lambda} \\ \\ \sqrt{\Lambda} & -\varphi_e(r)
\end{array}\right]
\left[\begin{array}{cc} \Psi_e(r) \\ \\ \Psi_s(r)
   \end{array}\right]\,,
\label{eq:msw}
\end{equation}
where
\begin{equation}
   \label{2} \varphi_e(r) = \frac{1}{4 E} \left( \pm
  2 \sqrt{2}\ G_F \left[
   N_e^-(r) - N_e^+(r) - \frac{N_n(r)}{2} \right] E - \delta m^2
   \cos{2\theta_v} \right)
\label{eq:potne}
\end{equation}
The upper sign is relevant for neutrino transformations; the lower one is
for antineutrinos.  In these equations
\begin{equation}
   \label{eq:offdia}
   \sqrt{\Lambda} = \frac{\delta m^2}{4 E}\sin{2\theta_v},
\end{equation}
$\delta m^2 \equiv m_2^2 - m_1^2$ is the vacuum mass-squared
splitting, $\theta_v$ is the vacuum mixing angle, $G_F$ is the Fermi
constant, and $N_e^-(r)$, $N_e^+(r)$, and $N_n(r)$ are the
total proper number densities of electrons, positrons,
and neutrons respectively in the
medium. Again, this form of the potential, $\varphi_e(r)$,
is valid only in the absence of
neutrino background effects.  The background effects will be explored in
detail in Section \ref{sec:back}, but we note that Eq. (\ref{eq:msw}) is
adequate for following $\nu$ flavor evolution well above the neutron star.

Eq. (\ref{eq:potne}) can be rewritten in
terms of a potential,
\begin{equation}
V(r) = 2 \sqrt{2} G_F \left[ N_e^-(r) - N_e^+(r) - \frac{N_n(r)}{2} \right].
\label{v-def}
\end{equation}
Then the on-diagonal term in the Hamiltonian becomes
\begin{equation}
  \varphi_e(r) = \frac{1}{4E}
                 \left( \pm V(r) E - \delta m^2 \cos 2\theta_v \right).
  \label{eq:subst-v}
\end{equation}
Neutrinos undergo a resonance and transform primarily when the on-diagonal
term is zero, that is, at the resonant energy
\begin{equation}
  E_{res}(r) = \pm\frac{\delta m^2 \cos 2\theta_v}{V(r)}.
  \label{eq:eres}
\end{equation}
We see that neutrinos undergo resonance when the potential is positive, and
antineutrinos undergo resonance when it is negative.  Finally, we rewrite
the potential in terms of the electron fraction,
\begin{equation}
   \label{eq:potye}
  V(r) = \frac{3 G_F \rho (r)}{2 \sqrt{2}m_N}
         \left( Y_e - \frac{1}{3} \right),
\end{equation}
where
\begin{equation}
   \label{eq:ye}
  Y_e (r)= \frac{N_e^-(r)-N_e^+(r) }{N_p(r)+N_n(r)}.
\end{equation}
Here $\rho(r)$ is the density of the matter at distance $r$ from the
protoneutron star, and $m_N$ is the mass of a nucleon
and $N_p(r)$ is the total proper proton number density.  We assume that
$N_p = N_e^- - N_e^+$ because of local electromagnetic charge neutrality.
Since the electron fraction can take on values between
zero and one, the potential can be either positive or negative.
Therefore, depending on the value of the electron fraction, either
neutrino or antineutrino transformation may be matter-enhanced.

We solve Eq. (\ref{eq:msw}) numerically for survival probabilities
of neutrinos as they pass through the matter above the protoneutron
star.  No approximations are employed in computing the survival
probabilities, for the adopted potential.
In the absence of oscillations and matter-enhanced flavor transformation,
the spectrum of neutrinos and  antineutrinos can be parameterized as
approximately Fermi-Dirac in character.  We
assume that no transformation has taken place within the protoneutron
star, so we begin with a full complement of each species of
active neutrino.
However, since neutrinos of different energies will have
different survival probabilities at each distance above the surface,
the distribution quickly departs from the Fermi-Dirac shape.  We use
representative initial neutrino distributions with temperatures of
${\rm T}_{\nu_e} = 3.22 \,
{\rm MeV}$ and ${\rm T}_{\bar{\nu}_e} = 4.5 \, {\rm MeV}$,
luminosities of $L_{\nu_e} = 1.08 \times 10^{51} {\rm ergs} \, {\rm
s}^{-1}$ and $L_{\bar{\nu}_e} = 1.3 \times 10^{51} {\rm ergs} \, {\rm
s}^{-1}$  and  effective chemical potentials of zero. Though different
numerical simulations of neutrino transport in the hot protoneutron star
core differ in their predictions of these spectral values, we note that
our adopted values serve to illustrate the general qualitative behavior
that would be expected if the simulations included this neutrino flavor
transformation physics.

In the neutrino driven wind, the neutrinos and antineutrinos
are the most important agents in determining the electron
fraction. However, near the surface of the protoneutron star there is
also a contribution from electrons and positrons:
\begin{equation}
\label{eq:ccf}
\nu_e + {\rm n}  \rightleftharpoons {\rm p}+ e^{-};
\end{equation}
\begin{equation}
\label{eq:ccbf}
\bar{\nu}_e + {\rm p} \rightleftharpoons {\rm n} + e^{+}.
\end{equation}
Close to the surface of the protoneutron star, these weak capture rates are
fast in comparison with the outflow timescale.  As we move far from the
surface, they become negligible in comparison with outflow.  Therefore the
weak capture rates are in the process of freezing out of steady state
(chemical) equilibrium.  We calculate the electron fraction by numerically
following the evolution of the number densities of neutrons and protons.

Our calculations are performed by tracking the evolution of mass elements
in the neutrino driven wind.  We use distance and density profiles from
analytic descriptions of the wind; $r \propto \exp(-t/\tau)$ and $\rho
\propto r^{-3}$ where $\tau$ is the outflow timescale \cite{qw}.  For
illustrative purposes, we use a timescale of $ \tau = 0.3 \, {\rm s}$ and
an entropy per baryon of $s = 100$ in units of Boltzmann's constant.  Close
to the surface of the protoneutron star, before the wind begins to operate,
we use the density profile of Wilson and Mayle \cite{wilson}.  Calculations
done with different timescales show the same qualitative features that we
present here \cite{us} and variations in the entropy are expected to have a
similar effect.

At each time step the distance and density are incremented.  All other
thermodynamic quantities, including the number densities of positrons and
electrons, are calculated from the entropy and density.  The mass fractions
of the neutrons, protons, alpha particles and heavy nuclei (A $>$ 40) are
calculated in NSE (Nuclear Statistical Equilibrium).  The weak rates are
then computed and the electron fraction is updated.  This new electron
fraction is used in the MSW equations to calculate new survival
probabilities for each neutrino energy bin.  We terminate calculations
for a particular mass element at the point when heavy nuclei begin to
form.  Since the outflow times considered in our calculations are
relatively short, $t \sim \tau
\lesssim 0.5 \, {\rm s}$, we have assumed that each mass element
experiences the same evolution as the previous ones.

The consequences of the feedback effect are illustrated in Figure
\ref{fig:efraction1}.  In this figure, the electron fraction is plotted
against distance as measured from the center of the protoneutron star.  The
upper curve shows the evolution of the electron fraction in the absence of
neutrino transformation, while the lower curve shows the evolution of the
electron fraction for mixing parameters of $\delta m^2 = 20 \, {\rm eV}^2$,
$\sin^2 2 \theta_{v} = 0.01$.  The initial rise in the electron
fraction is due to Pauli unblocking of neutrino capture on neutrons as the
density rapidly decreases at the edge of the protoneutron star.  At such
high density we do not include feedback effects; the validity of this
approximation is discussed below.  We begin the feedback effects at about
the distance where the wind solution begins to dominate the hydrodynamic
flow.  In the case of no transformation the electron fraction shows a
slight drop, due to the decreasing importance of electron and positron
capture, and then a slight rise due to the alpha effect.

In the presence of mixing, however, the situation is quite different.  Low
energy electron neutrinos begin to transform slightly above the surface of
the protoneutron star, where the potential is large (Eq.~\ref{eq:eres}).
This transformation decreases the rate of neutrino capture on neutrons,
which lowers the electron fraction, and therefore the potential.  Thus
neutrinos of higher and higher energy begin to transform.  Eventually the
electron fraction drops below 1/3, so the potential becomes negative, and
the highest energy electron antineutrinos begin to transform. This slows
down the drop in $Y_e$, but does not halt it entirely, since low energy
electron antineutrinos are still present.  As the electron fraction
continues to fall, the magnitude of $V$ increases, so lower energy electron
antineutrinos transform.  Meanwhile, the mass density is falling, causing
the potential to reach a minimum and, eventually, return toward zero.
The minimum means that only antineutrinos above a certain energy will
undergo resonance, and since the potential goes back to zero, all these
neutrinos (originally active antineutrinos) are re-converted from sterile
to active species.  By the time alpha particles form, there are many
electron antineutrinos present but few electron neutrinos. Therefore,
there is no alpha effect.  The resulting electron fraction is so low
$\sim 0.1$, that conditions become favorable for the r-process. There is
no alpha effect because the $\nu_e$'s that would have captured on
neutrons during the epoch of alpha particle formation are, for the right
neutrino mass/mixing parameters,  converted to
sterile species in large measure at the alpha formation epoch.  A study of
many different neutrino driven wind conditions, shows that for a
timescale of 0.3~s and an entropy per baryon of $\sim 100$, the electron
fraction must be below 0.18 in order to produce the requisite neutron to
seed ratio
\cite{meyer97}.

An analysis in Ref.\ \cite{nunokawa} assumed that $Y_e(r)$ was equal to
its equilibrium value, as set from the weak capture rates
\cite[Eq.~3.14]{us}. Ref. \cite{nunokawa} neglected electron and
positron capture in finding the equilibrium value of of $Y_e(r)$, and
did not take into account alpha particle formation and did not take
account of the feedback of the outflow rate on neutrino capture
processes and neutrino flavor transformation.  Similarly,
Ref. \cite{peres} did not include this physics.  
Without these effects, the
system finds a fixed point at
$Y_e = 1/3$; neutrino conversions bring the electron fraction to this
value. Once there, the neutrino-matter forward scattering-induced
potentials vanish and there is no further neutrino flavor transformation
in the channels $\nu_e\rightleftharpoons\nu_s$ and
$\bar\nu_e\rightleftharpoons\bar\nu_s$.
In Fig. \ref{fig:equ}, we have duplicated this result by making the
assumptions outlined above.  The figure also shows the evolution of $Y_e$
when we take account of alpha formation, but keep the other
simplifications as above.  In this case, $Y_e$ goes to 1/3 as before, but
then the alpha effect drives it to 1/2 eventually.  These results hold
for a broad range of neutrino mixing parameters.

Except in Fig. \ref{fig:equ} we have carefully tracked the actual
value of $Y_e$ as distinct from its equilibrium value; we have
also taken account of electron and positron captures and alpha
formation. In this case, $Y_e$ lags behind its equilibrium value.
As a result, $Y_e$ remains greater than 1/3, even when the equilibrium
electron fraction ${(Y_e)}_{\rm eq}$ drops below that value.
Therefore, neutrinos continue to transform, and eventually drive the system
to low electron fraction values, $Y_e < 1/3$.

We investigate a range of mixing parameters in Figure \ref{fig:efraction2}.
This contour plot shows the electron fraction, measured at the point where
heavy nuclei begin to form, for various mixing parameters. Here we employ
the same neutrino driven wind model used in Figure \ref{fig:efraction1}.
In the bottom left corner of the plot, the solution is approaching the
case without neutrino transformation.  In the middle of the plot, the
transformations of neutrinos produce a very neutron rich environment.

In the upper right part of the plot, the electron neutrinos are undergoing
an additional resonance, at smaller distance than those described above, and
in fact close to the neutron star surface where the density scale height is
small. This resonance occurs at the first place where the electron fraction
passes close to 1/3, at very high density.  To correctly model this range of
mixing parameters it is necessary to include feedback in the very dense
region near the neutron star surface.
 
In addition, because this resonance is
so close to the neutrinosphere, some neutrinos travel through it along
extremely nonradial paths.  In general neutrinos which travel nonradially
transform differently from those which travel radially, but for most of our
parameter space, the effects are small.  The results presented here include
only radial paths. We have checked them against another (very slow)
calculation, which includes nonradial effects.  The difference is negligible
except when neutrinos or antineutrinos transform at this inmost resonance.

We have drawn a shaded area on the contour
plot.  In this shaded area the inclusion of nonradial neutrino paths
would alter the results by more than 10\%.  We show this also in Fig.
\ref{fig:nonradial}. Here the difference in survival probabilities of
electron neutrinos at 11 km is plotted.  The same wind parameters are
used here as in Fig. \ref{fig:efraction2}.

This shaded region of Fig. \ref{fig:efraction2} (and also
the unshaded region in the upper right corner above it)
  would also require us to take
account of feedback effects in the steep density gradient region near the
neutrino sphere.  We have not included a full nonradial calculation of the
results there.  These regions of the plot require
 further investigation before
definitive conclusions about the evolution of the electron fraction can be
drawn.  The work of Ref. \cite{mitesh} has discussed the difficulties
inherent in treating the steep region. 
We emphasize, however, that the rest of our parameter space presents a
favorable environment for the
$r$-process.

\section{Detection}
\label{sec:detect}

In this section we study the supernova neutrino signal that would be
seen at SNO or SuperKamiokande after active-sterile transformation.
We consider the $\nu_e$,
and $\bar{\nu}_e$ parameters used in section II and
additional individual neutrino parameters of
$T_{\nu_\mu,\bar{\nu}_\mu,\nu_\tau,\bar{\nu}_\tau}= 6 \, {\rm MeV}$ and
$L_{\nu_\mu,\bar{\nu}_\mu,\nu_\tau,\bar{\nu}_\tau} = 1 \times 10^{51}
{\rm ergs} \, {\rm s}^{-1}$ .

A galactic supernova is expected to produce a large electron
antineutrino signal in SuperK.  In fact, a supernova at 10 kiloparsecs is
estimated to see about 8000 events from
$\nu_e + n \rightarrow e^- + p $ \cite{superK}.
In principle, there will be an additional 300 events
from neutrino scattering on electrons.  Since both charged and neutral
current processes contribute to this type of scattering, all types of
neutrinos, except sterile will contribute.

SNO will see an estimated 500 events  from neutral current
break-up of the deuteron, $ \nu_x + d \rightarrow n + p$ and an
additional 150 for each of the charged
current processes, $\nu_e + d \rightarrow p + p + e^{-}$ and
$\bar{\nu}_e + d \rightarrow n + n + e^{-}$ \cite{SNO}.
These break-up reactions may be
distinguished and tagged by the presence or absence of the emitted neutron(s)
and the Cerenkov light from the electron or positron.

\subsection{Active-sterile transformation alone}

Figure \ref{fig:sno1} shows the ratio of the expected charged current electron
neutrino-induced event rates in SNO for the case with active-sterile
transformation to the case without such transformation.  In the region which
produced the lowest electron fraction in Figure \ref{fig:efraction2}, we find
that the charged current event rate is suppressed by 90\%.  In the lower left
hand corner of the plot, the ratio approaches one as the solution
asymptotes to the case of no transformation.  In Figs. \ref{fig:sno1} -
\ref{fig:elscat2}, the shaded region shows where the neutron to seed
ratio should be greater than 100 (i.e. favorable for the r-process) 
for the given wind parameters.  

Figure \ref{fig:elscat1} shows the ratio of the expected electron
neutrino-induced scattering events from all processes at Superkamiokande
for the case with active-sterile
transformation to the case without such transformation. Because the
charged current electron neutrino scattering rate is so much larger than the
neutral current one, the absence of electron neutrinos has a significant
$\sim 40$\% effect. This may be surprising since the
$\nu_\mu$,$\bar{\nu}_\mu$,$\nu_\tau$, and $\bar{\nu}_\tau$ have two and a
half times the energy of the
$\nu_e$ on average and the cross section is approximately linear in
neutrino energy.  However, the luminosities of the species are roughly
the same, so the number flux of the high energy neutrinos is two and
half times less than the number flux of the lower energy neutrinos.
Therefore most of the small energy dependence in the rate stems from
the 5 MeV detection threshold in SuperK.
 
These figures demonstrate that because of the deficit of electron neutrinos
produced by the active-sterile solution, the SNO charged current rate will
be greatly reduced from the expected event rate.  However, the SuperK electron
(antineutrino) capture rate will be for the most part unchanged.  Note that
our model is, therefore, unconstrained by data from supernova 1987A.

The calculations of detector signals have assumed no
other type of neutrino mixing until this point.
However, both the atmospheric
neutrino mixing solution and the solar neutrino solution will cause
additional oscillations.  The effect of $3 \, \times \,  3$
 active only mixing on
a supernova neutrino signal has been discussed in the literature
(ref. \cite{fhm,dighe}).  The vacuum mixing of $\nu_\mu
\leftrightarrow \nu_\tau$ which may solve the atmospheric neutrino
problem will cause $\nu_\mu \leftrightarrow \nu_\tau$ mixing
in supernova neutrinos. Since
the $\nu_\mu$ neutrinos and the $\nu_\tau$ neutrinos have the same
energies and luminosities, this mixing will not directly
cause changes in the detector signal.  In the following discussion,
we assume $\sin^2 2 \theta_{atm} \approx \sin^2 2 \theta_{23} = 1$.
 
\subsection{SMA and active-sterile transformation}

We also consider the situation  where $\nu_e
\leftrightarrow \nu_\mu$ MSW is the solution to the solar neutrino
problem.  The small mixing angle (SMA) parameters,
$\delta m^2_{solar} = 5 \times 10^{-6} \, {\rm eV}^2$ and $\sin^ 2
2 \theta_{solar} = \sin^2 2 \theta_{12} \approx 10^{-2}$
will cause partial conversion in the hydrogen envelope of
the supernova progenitor star. 
The exact survival probabilities will depend on the
electron density scale height:
\begin{equation}
R_s = -  { N_e \over dN_e/dr}.
\end{equation}
We use densities and distances from Ref. \cite{woosley} to estimate $R_s
\approx 10^{10} {\rm cm}$. We assume here that the density and
composition will be largely unaffected by the core collapse event at the
time that the neutrinos move through it.  
For 30 MeV neutrinos, near-total
conversion (electron neutrino survival probability less than 1\%) requires
the density profile to be much shallower: $R_s > 8 \times10^{10}$ cm.  For
no conversion (electron neutrino survival probability more than 99\%), the
density profile must be much steeper: $R_s < 1 \times 10^{8}$ cm.
Therefore it is very likely that partial conversion of the electron neutrinos
will take place, even given a more detailed model of the hydrogen envelope.

Using a fit to the points given in ref. \cite{woosley}, we obtain ratios
for the SNO charged current electron neutrino scattering rate (Figure
\ref{fig:sno2}) and the SuperK electron neutrino rate (Figure
\ref{fig:elscat2}), with both the active-sterile mixing and the
additional envelope conversion.  Again, we show results in terms of
ratios: the expected detector supernova neutrino-induced event rates with two
types of mixing to the expected event rates in the case with no neutrino
flavor conversion.  The SuperK electron antineutrino capture rate is
the same as in Figure 4, since no electron antineutrino conversion occurs in
the envelope.

Figure \ref{fig:sno2} shows that the partial conversion of muon neutrinos to
electron neutrinos, after the initial conversion of electron neutrinos
to steriles, produces a ratio of order one.
Figure \ref{fig:elscat2}
  shows that electron scattering
ratio of rates at SuperK with conversion.  The partial conversion of the muon
neutrinos more or less reproduces the case with no transformation whatsoever.

\subsection{LMA and active sterile transformation}

In table \ref{tab:osc} we show a few of the possible
outcomes for final electron neutrino spectrum, as a result of the
active-active mixings.  The relevant parameters are  $\theta_{12}$,
$\theta_{13}$ and the associated $\delta m^2$s.  The case of maximal
mixing for the atmospheric neutrinos and the large angle solution to the
solar neutrino problem, with $\sin^2 2 \theta_{12} \approx
\sin^2 2\theta_{solar}  \sim 0.8$ is shown.
Here we implicitly assume a normal (as opposed to an inverted) neutrino mass
hierarchy.

As in the case of small angle mixing, the Large Mixing Angle (LMA)
solar parameters will cause neutrino flavor conversion in the supernova at a
density comparable to where resonant neutrino flavor conversion takes place in
the sun.  In the supernova progenitor, this resonance will be located in
the helium/hydrogen envelope for typical neutrino energies.   If the
supernova progenitor density profile is fairly smooth, then there will likely
be an adiabatic level crossing at this location, so that
the $\nu_e$
will wind up with an energy spectrum which would be measured as
 $\sin^2 \theta_{solar}$ of the original
$\nu_e$ spectrum and $\cos^2 \theta_{solar}$ of the originally ``hotter'' 
$\nu_\mu$ and $\nu_\tau$ type spectrum.  
The \lq\lq hopping probability\rq\rq\ at
resonance will be much smaller for the large angle solution than in the case
of the SMA.
 For the LMA solution, even if the resonance is
completely nonadiabatic so  that there is no level crossing 
at resonance,  the electron neutrinos will still become more energetic on
account of their large vacuum mixing with mu and tau neutrinos. 

With the LMA included,
the difference between the active-sterile neutrino conversion case considered
here and the case with no steriles is that part of the $\nu_e$ spectrum
simply  disappears at some point  during the deleptonization \lq\lq
cooling\rq\rq\ epoch of the proto-neutron star.   At early times, when the
density profile is too steep for much active-sterile transformation to occur,
the ``cold'' part (or the part that was originally $\nu_e$) of the
spectrum will be present, and at late times, when the transformation
begins and the r-process takes place,
this part will disappear. We show this pictorially in
Fig. \ref{fig:earlylate}.

In Fig. \ref{fig:earlylate} the bottom panel shows what happens to the neutrino
spectrum if $\nu_e \leftrightarrow \nu_s$
resonant neutrino flavor conversion is significant
  and no subsequent vacuum mixing or envelope conversion takes place.
The solid curve corresponds to the early time, when the
density profile is too steep for much flavor transformation to occur and
the dashed curve corresponds to the late time when the evolution is
adiabatic.

The top panel of Fig. \ref{fig:earlylate}
shows the case where the
LMA parameters provide the solution to the solar neutrino problem.  We assume
completely adiabatic
transformation in the supernova progenitor envelope in this case.   In this
panel, the original $\nu_e$ spectrum becomes mixed with the $\nu_\mu$ and
$\nu_\tau$, through the  solar parameters $\sin^2 2 \theta_{solar} \def \sin^2
2\theta_{12} \approx 0.8$. This is evident in the figure: note, for 
example, the
longer tail on the  distribution function.  The early time (solid line) case,
where the $
\nu_e
\leftrightarrow \nu_s$ transformation is inoperative, has many more counts
at low energy than does the late time scenario.  From this curve it is
evident that measuring the low energy
part of the neutrino spectrum would be most useful for determining whether
large scale active-sterile transformation occurs.  A dramatic change in the
number of low energy neutrinos relative to the number of high energy neutrinos
would be an indication that the type
of active-sterile transformation discussed here was taking place.

\subsection{The third active-active mixing angle}

An interesting case occurs if the unknown parameters $\theta_{13}$ and
$\delta m_{13}^2$ are such that a second resonance occurs at a density
between the $\nu_e \leftrightarrow \nu_s$ transformation density and the
resonance region associated with
$\delta m^2_{12}$.  In the case of complete neutrino flavor transformation, the
$\nu_e$'s are almost completely transformed into
$\nu_{\mu,\tau}$ and vice versa.  In this case,
neutrinos which were originally electron flavor when they left the neutrino
sphere are almost  completely coincident with the $\nu_3$ state.
This state evolves separately and does not encounter subsequent mixing
with the other flavors \cite{bf,cfq}.
In this special case of complete transformation
at the
$\delta m^2_{13}$ resonance, it would be more difficult
  to recognize the effects of
a $\nu_e \leftrightarrow \nu_s$ transformation at a higher
density.

\subsection{Detection Summary}
If only active-sterile transformation takes place, there
would be a dramatic deficit in the expected charged current signal in SNO.
However, with $\nu_e \leftrightarrow \nu_{\mu,\tau}$ transformations 
included as
well,  whether the detected signal is distinguishable from the case with no
steriles depends on the oscillation parameters and the density and electron
fraction profiles in the supernova.  One signal of the active-sterile 
mixing and
the LMA would be a decrease in the number of low energy neutrinos with time.

We note that we have not considered in this section the mixing
of the sterile neutrino with the other (mu and tau) active neutrino species.
Such mixing opens a large new range of parameter space.  Depend on the
$\delta m^2$s and mixing angles, this additional unknown
mixing could alter the
results presented in this section.  Finally, density fluctuations,
both near the protoneutron star and in the progenitor envelope
may also change these results \cite{loreti}.  

\section{Background}
\label{sec:back}

In this section we consider the effects of the neutrino background
potential.  The neutrino-neutrino forward scattering-induced \lq\lq
background\rq\rq\ potential adds an important new twist to resonant neutrino
flavor conversion, rendering the problem severely nonlinear. This potential can
be written
\begin{equation}
V_\nu = 2\sqrt{2}G_F N_\nu.
\label{eq:nunu-v-def}
\end{equation}
Here $N_\nu$ is the {\bf effective} (because a neutrino's individual
contribution depends on the angle it makes with the \lq\lq 
test\rq\rq\ neutrino)
net (neutrinos minus antineutrinos) neutrino number density.  In the case of
active-sterile mixing, only the flavor basis on-diagonal terms entering
into Eq. 
(\ref{eq:msw}) are non-vanishing, although in the general case of active-active
mixing, both on- and off-diagonal matrix elements of the neutrino-neutrino
forward scattering potential in the flavor basis could be nonzero
\cite{pantaleone:offd,sigl-raffelt:kinetic}.   We only need to take account of
electron neutrinos forward-scattering on active flavors, since we 
assume that the
matrix elements for electron neutrinos to forward-scatter off steriles are
negligible, as are the matrix elements for sterile neutrino forward-scattering
off other sterile species
\cite{sigl-raffelt:kinetic}.  See
Ref.~\cite{qf:background} for a discussion of the background effect in
active-active neutrino mixing.

The effective neutrino number density is written as
\begin{equation}
N_\nu = \int d^3 q \left( 1 - { p \cdot q \over |p| |q|} \right)
(\rho_{q,aa} - \bar{\rho}_{q,aa}).
\end{equation}
Here $q$ is the momentum of the background neutrino, $p$ is the
momentum of the test neutrino, $\rho_{q,aa}$ is the density matrix
element for neutrino-neutrino forward scattering and $\bar{\rho}_{q,aa}$ is the
corresponding density matrix element for the antineutrinos.  In these
expressions for the density matrix elements, the subscript $aa$ denotes the
flavor-diagonal matrix element, so that $a=e$,$s$, electron or sterile flavor,
respectively. We write the density matrix elements as
\begin{equation}
\rho_{q,aa} = N(r,R_\nu,E_q) P(r,E_q,\psi)
\end{equation}
where $E_q$ is the energy of a neutrino with momentum $q$.
$N(r,R_\nu,E_q)$ is the number density of neutrinos
emitted into a pencil of direction and energy interval $d^3q$, corresponding
to the neutrino number flux (divided by the speed of light $c$) at a 
given energy
in the pencil of directions centered on the angle
$\psi$. Here
$\psi$ is the angle of emission measured from the normal to the
neutrinosphere.
The function
$P(r,E_q,\psi)$ is the survival probability as a function of given
energy and emission angle.

We find that for a test neutrino traveling at an angle $\alpha_t$
to the radial,
then $N_\nu$ can be written as
\begin{equation}
   N_\nu (r,\alpha_t) =
      2\pi N_\nu^0
      \int_0^\infty dE_\nu f_\nu(E)
                    \left( Q_0(r,E_\nu) - \cos\alpha_t Q_1(r,E_\nu) \right).
\label{eq:nveff}
\end{equation}
where
\label{eqs:ang-moments}
\begin{eqnarray}
  Q_0(r,E_\nu) & = & \int_\mu^1 d\cos\alpha_b \, P(r,E,\alpha_b), \\
  Q_1(r,E_\nu) & = & \int_\mu^1 d\cos\alpha_b \, P(r,E,\alpha_b) \cos\alpha_b,
\end{eqnarray}
Here, $\mu$ is defined as $\mu = \sqrt{1 - (R_\nu/r)^2}$, $\alpha_b$ is
the angle of the background neutrino with respect to the radial, and
$N_\nu^0$ is the number density of the neutrinos at the neutrino sphere.

 From these expressions it is clear that when including background effects,
the potential depends on the survival probability of neutrinos which
travel at different angles and take different paths.

In calculations with the neutrino background potential included,
it is most convenient to evolve bilinears of the
wave functions.  For much of the evolution, it is actually more efficient
to change to the matter basis and evolve with the angle-phase parameterization
~\cite{pantaleone:isotropic}.

To compute the neutrino potential at each time step, we must find the
active-active neutrino survival probability at each neutrino energy.  In
the matter basis, the survival probability depends on the matter angle; but
in turn, the matter angle depends on the survival probability through
$V_\nu$.  For a given set of matter states, then, we must find a $V_\nu$
which gives a consistent set of survival probabilities and matter angles.
Through most of the evolution, there is a unique neutrino potential which
is consistent with the matter states.  When large numbers of neutrinos
begin to transform, though, there can be multiple consistent potentials.
In this range we eliminate the ambiguity by changing to the matter basis;
we change back when there is little transformation.

\subsection{Radial Results}

To simplify the problem, we have assumed that the survival probability does
not depend on the neutrino's direction.  The next section discusses some
implications of relaxing this assumption.  We do not simply use
$\alpha_t = 0$, since this underestimates the  potential.
Instead, we take the survival probability to be independent of direction:
\begin{equation}
  Q_0(E_\nu) = P(E_\nu)(1-\mu)
\end{equation}
and
\begin{equation}
  Q_1(E_\nu) = P(E_\nu)(1-\mu^2)/2.
\end{equation}
The neutrino potential is then given as
\begin{equation}
   N_\nu^{\mathrm{avg}} = 2 \pi N_\nu^0 (1-\mu)^2 \left( \frac{3+\mu}{4} \right)
                          \int_0^\infty dE_\nu f_\nu(E_\nu) P(E_\nu).
\label{eq:navg}
\end{equation}
To leading order, the geometric factor $(1-\mu)^2 (3+\mu)/4$ is equal to
$(R_\nu/r)^4/4$.  This is double the leading-order potential we would find
by taking $\cos\alpha_t = 1$.
 
A sample run using this potential is shown in Figure \ref{fig:back}. In this
figure we reduce the neutrino luminosities by a factor of ten from the
calculations in the previous sections.  We do this to make the calculation
more tractable.  However, we also note that these luminosities are not
unrealistic at very late times, where the r-process elements are likely
still being made ~\cite{cfq}.

There are several effects stemming from the additional neutrino potential
provided by the neutrino forward scattering-induced background.  One is that
there is somewhat more transformation in the innermost resonance 
relative to the
no neutrino background case. The neutrino background potential changes more
slowly than the external potential provided by neutrino forward scattering on
the electrons: the result is to make neutrino amplitude evolution through the
resonance more adiabatic. Smaller values of
$\delta m^2 \sin^2 2\theta_v$ will reduce the adiabaticity, however. Overall,
though the increase in adiabaticity afforded by a significant neutrino
background potential implies more transformation of neutrino flavors at
resonance and does suggest that the effect of the neutrino background will be
to extend the epoch of significant neutrino flavor transformation out to
relatively smaller luminosities and, hence, later times than would be the case
with a purely electron-driven neutrino flavor transformation potential in
operation.

Another effect of a significant neutrino background arises in the wind
epoch/region. As in the purely electron-driven neutrino flavor
transformation case, but not so monotonically, the resonance energy will tend to
sweep from low energy to high energy through the
$\nu_e$ energy distribution.  As
$\nu_e$'s begin to transform,
$V_\nu$ decreases.  Its decrease becomes faster than that of the external
potential; as $V_\nu$ decreases, the higher-energy resonances become
nonadiabatic.  Therefore, with a significant neutrino background we have less
flavor transformation than in the no-background case.  Finally, $V$ 
becomes small
enough that
$E_{res}$ is larger than the maximum $\nu_e$ energy; at this point, 
we have swept
through the entire
$\nu_e$ population.

After $E_{res}$ has passed through the $\nu_e$ distribution, 
significant neutrino
flavor transformation mostly ceases.  Except for the slow decrease of $Y_e$ and
$\rho$, there is nothing more to change the potential, and $\bar{\nu}_e$
transformation is minimal; even though the potential becomes negative, the
$\bar{\nu}_e$ resonant energy is always above the $\bar{\nu}_e$ distribution.
The
survival probabilities, then, are essentially fixed.  $Y_e$ continues to
fall towards its equilibrium value, and in this example
(Fig. \ref{fig:back}), ends up fairly
low.  However, even in this example there is a non-negligible alpha effect,
visible not as an upturn but a flattening of the $Y_e$ profile.  At higher
luminosities, fewer neutrinos need to transform in order for $V_\nu$ to
balance the external potential.  The increased $\bar{\nu}_e$ population then
blocks the lowering of $Y_e$ and causes a larger alpha effect.  A sizable
radial background effect thus blocks our mechanism, or at least changes the
optimal $\delta m^2$ and $\sin^2 2 \theta_v$ at which it would occur.

The larger the luminosity, of course,
the fewer the number of neutrinos which must transform in order to drive the
potential
$V$ to zero. However, large luminosity means $V_\nu$ will fall faster once
$\nu_e$'s start to transform to sterile species, and therefore the resonances
will tend to be less adiabatic.  The situation is, in some ways, analogous to
the case where
$Y_e$ is set equal to its equilibrium value.  In that case, as here with a
significant neutrino background, there is immediate feedback between survival
probabilities and the MSW potential $V$; in contrast, feedback 
without a neutrino
background is substantially delayed.  Because there is no delay in the feedback
with a significant neutrino background, once $V$ is driven near zero, 
there is no
more reason for it to change---the system has found an equilibrium.  An
independent calculation of the active-sterile background effect~\cite{mitesh}
finds similar effects on
$Y_e$.

\subsection{Nonradial Speculations}

When we use a radial treatment for the background, neutrinos evolve until
$V=0$, then stop.  Prior treatments of background have also used this
approximation.  However, the dominance of the fixed point at $V=0$ makes a
radial treatment unsuitable in our situation, because neutrinos coming in
at different angles will see different potentials.  Therefore, a full
treatment will not have the same fixed point we see in the radial
approximation.

A complete treatment of the background, including nonradial neutrino paths,
may be possible by an extension of the treatment described above.  In
addition to dividing neutrinos into bins of energy, one would bin them
according to their emission angle $\psi$.  Because of the spherical
symmetry, neutrinos with the same $\psi$ will encounter the same potential
(including neutrino potential), and have the same survival probability.
A full nonradial treatment will likely give substantially different results
than the radial one.

\section{Conclusions}
\label{sec:concl}

We have considered the effect of  active-sterile neutrino transformation
on the r-process in the neutrino driven wind environment.
Preliminary calculations show that it is possible to
drive the material sufficiently
neutron-rich that rapid neutron capture is possible.

The effect on the neutrino signal from a nearby supernova depends very
much on the additional, active-active mixing parameters.  In the absence of
this additional mixing, the effect on the neutrino signal would be dramatic.
If the r-process in the neutrino driven wind is made possible by active-sterile
transformation, then their would be a dramatic lack of electron neutrinos
coming from the supernova at late times, which would correspond to a reduction
in the neutrino signal of $\sim 90\%$.

However, recent data from
Superkamiokande and SNO indicates that the additional active-active
transformations probably do take place.  The most relevant for the
discussion here are the transformations involving $\nu_e$.  In the case
of the SMA, we estimate about a 50\% tranformation for $\nu_e$ with
$\nu_{\mu,\tau}$.  In the case of the LMA, depending on the
exact value of $\sin^2 2 \theta_{solar}$, more than 50\% of the electron
neutrinos are likely to transformation to $\nu_{\mu,\tau}$ and vice-versa.
In either case, the signal of active-sterile neutrino transformation
is more difficult to identify, but is likely to marked by a decrease of
low energy neutrinos at late time.

We have also discussed background effects.  We have done calculations
for reduced luminosity and with radial neutrinos. The latter
  is the approximation
that all neutrinos see the same potential, regardless of their path.
We find
that the background effects tend to prevent the electron fraction
from dropping to values as low as in the nonbackground case.  This
implies that the effect is most successful for the producing the r-process
elements at late times, when the luminosity is low.  A detailed
investigation of neutrino background effects, 
using different methods and a different range of
luminosities has been undertaken recently in Ref. \cite{mitesh}.  
Finally, we note that nonradial background effects, which take
into account the different potentials seen by neutrinos traveling on different
paths, is  likely to be quite important, and we speculate that this may
significantly alter the results from the radial case.

\acknowledgements

This work was supported in part by the U.S. National Science
Foundation Grants No.\ PHY-0070161  at the University of Wisconsin and
No. PHY-0099499 at UCSD, and in part by the University of Wisconsin Research
Committee with funds granted by the Wisconsin Alumni Research
Foundation.  We thank the ECT* for their hospitality during part of the 
completion of this work.

\newpage

\begin{table}
\begin{tabular}{|c|c|c|l|c|c|c|c|}
\hline\hline
& \\
Complete $\nu_e \leftrightarrow \nu_s$ transformation  and
&
$\nu_e$ spectrum \\ \hline\hline
   $\theta_{13}$,$\delta m^2_{13}$ no conversion  & \\
  $\theta_{12}$,$\delta m^2_{12}$ adiabatic level crossing
&  3/4 ``hot'' + 1/4 ``nothing'' \\ \hline
  $\theta_{13}$,$\delta m^2_{13}$ no conversion    & \\
  $\theta_{12}$,$\delta m^2_{12}$ nonadiabatic level crossing &
  1/4 ``hot'' + 3/4 ``nothing'' \\ \hline
   $\theta_{13}$,$\delta m^2_{13}$ complete conversion & \\
  $\theta_{12}$,$\delta m^2_{12}$ adiabatic level crossing
&   all ``hot''   \\ \hline
   $\theta_{13}$,$\delta m^2_{13}$ complete conversion   &  \\
  $\theta_{12}$,$\delta m^2_{12}$  nonadiabatic level crossing
&  all ``hot''    \\ \hline
\hline
& \\
No steriles and
& $\nu_e$ spectrum \\ \hline\hline
$\theta_{13}$,$\delta m^2_{13}$ no conversion & \\
$\theta_{12}$,$\delta m^2_{12}$ adiabatic level crossing &
3/4 ``hot'' + 1/4 ``cold''   \\ \hline
$\theta_{13}$,$\delta m^2_{13}$ no conversion & \\
$\theta_{12}$,$\delta m^2_{12}$ nonadiabatic level crossing &
1/4 ``hot'' + 3/4 ``cold''   \\ \hline
$\theta_{13}$,$\delta m^2_{13}$ complete conversion & \\
$\theta_{12}$,$\delta m^2_{12}$ adiabatic level crossing &
all ``hot'' \\ \hline
$\theta_{13}$,$\delta m^2_{13}$ complete conversion & \\
$\theta_{12}$,$\delta m^2_{12}$ nonadiabatic level crossing &
all ``hot''    \\ \hline

\end{tabular}
\vspace*{0.5cm}
\caption{
\label{tab:osc}
Examples of various scenarios of additional neutrino oscillations in
the envelope.  The $\nu_e$ spectrum in the second column is the
spectrum of the neutrinos as they arrive at the earth.  We assume that
the $\nu_\mu, \bar{\nu}_\mu, \nu_\tau, \bar{\nu}_\tau$ neutrinos
have spectra with higher average energy (``hot''), then the electron
neutrino (``cold'').
We have associated the angle $\sin^2 2 \theta_{12}$
with $\sin^2 2 \theta_{solar} \sim 0.8$ and the angle
$\sin^2 2 \theta_{23}$ with
$\sin^2 2 \theta_{atmospheric} \sim 1$.
The angle $\theta_{13}$ is unknown and limited only by
the reactor neutrino data.
In this table we have neglected the possible consequences of mixing between
the sterile and the other active flavors.  In the case where there is
complete transformation in the $\delta m^2_{13}$ resonance,
there is no observable effect on the $\nu_e$ spectrum from the subsequent
$\theta_{12}, \delta m^2_{12}$ mixing.   If there is no transformation in
the $\theta_{13}$ channel, then the $\nu_e \leftrightarrow \nu_s$
oscillation solution may be detected as a loss of the low energy neutrinos,
see Fig. \ref{fig:earlylate}.}
\end{table}

\newpage
\clearpage

\begin{figure}
\centerline{\includegraphics[angle=0,width=14cm]{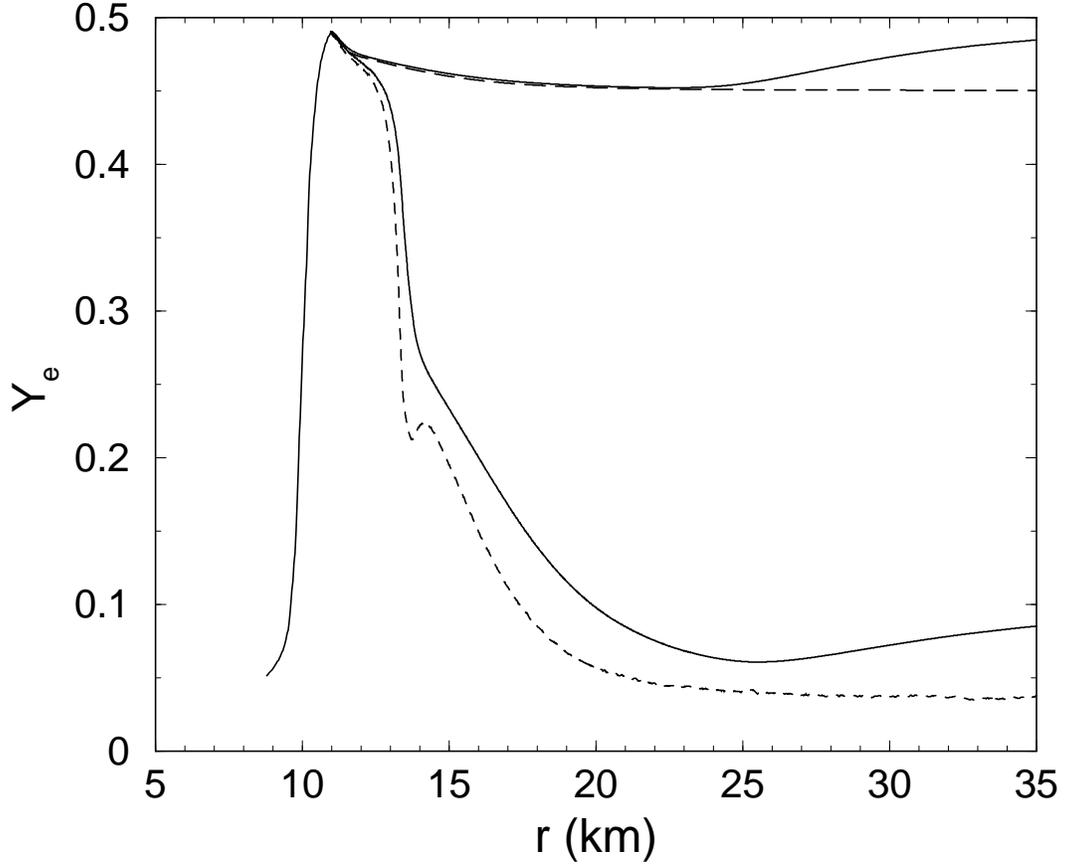}}
\caption{The electron fraction is plotted against distance from the
center of the neutron star.  The upper lines shows the evolution with no
transformation.  The lower lines shows the evolution of active-sterile mixing
parameters of $\sin^2 \theta_v = 0.01$ and $\delta m^2 = 20 \, {\rm eV}^2$. 
The dashed lines show the equilibrium electron fraction while the
solid lines show the actual electron fraction. 
}
\label{fig:efraction1}
\end{figure}

\begin{figure}
\centerline{\includegraphics[angle=0,width=14cm]{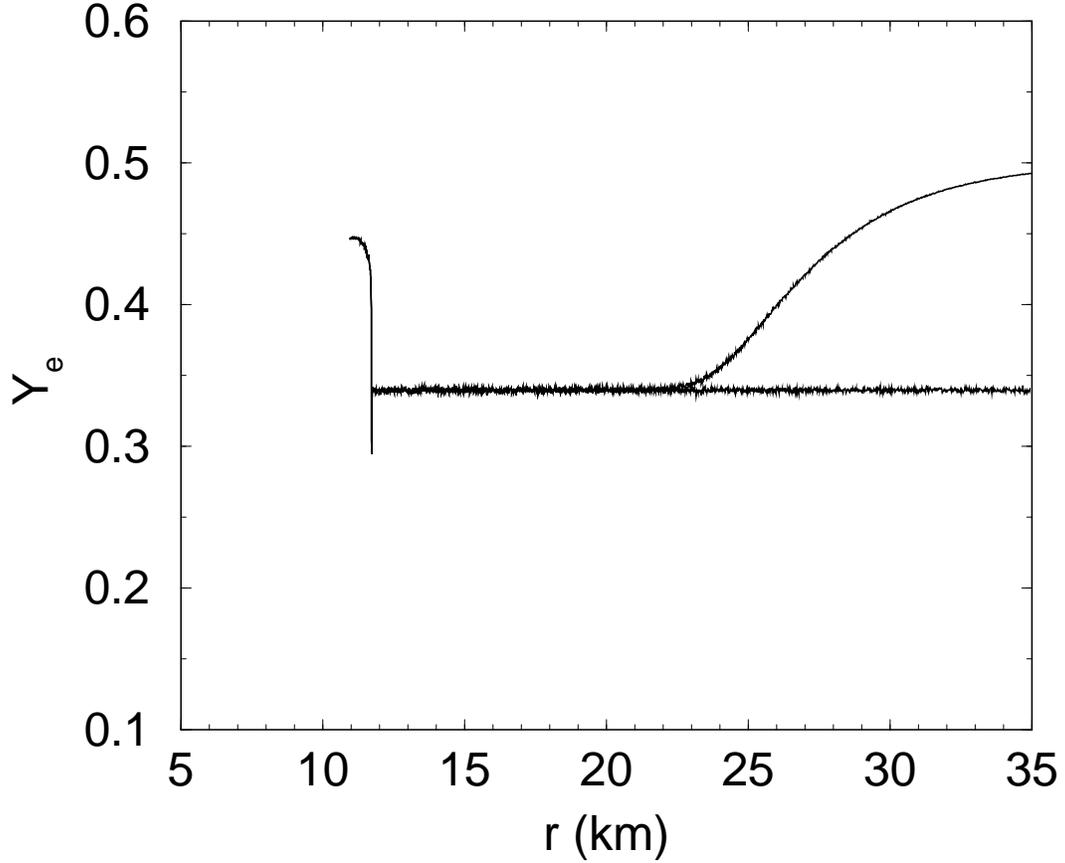}}
\vspace*{1cm}
\caption{The electron fraction is plotted as in Fig. \ref{fig:equ},
under the same neutrino mixing parameters;  however, we have assumed that
the electron fraction goes immediately to its equilbrium value, as
set by the neutrino capture rates.  In this figure only, we have neglected
electron and positron captures.  In the lower line, we have further
neglected the formation of alpha particles.  The upper line shows the effect
of including alpha particles.  The brief dip below 1/3 is an artifact of
the code's finite step size.
\label{fig:equ}}
\end{figure}

\begin{figure}
\centerline{\includegraphics[width=14cm]{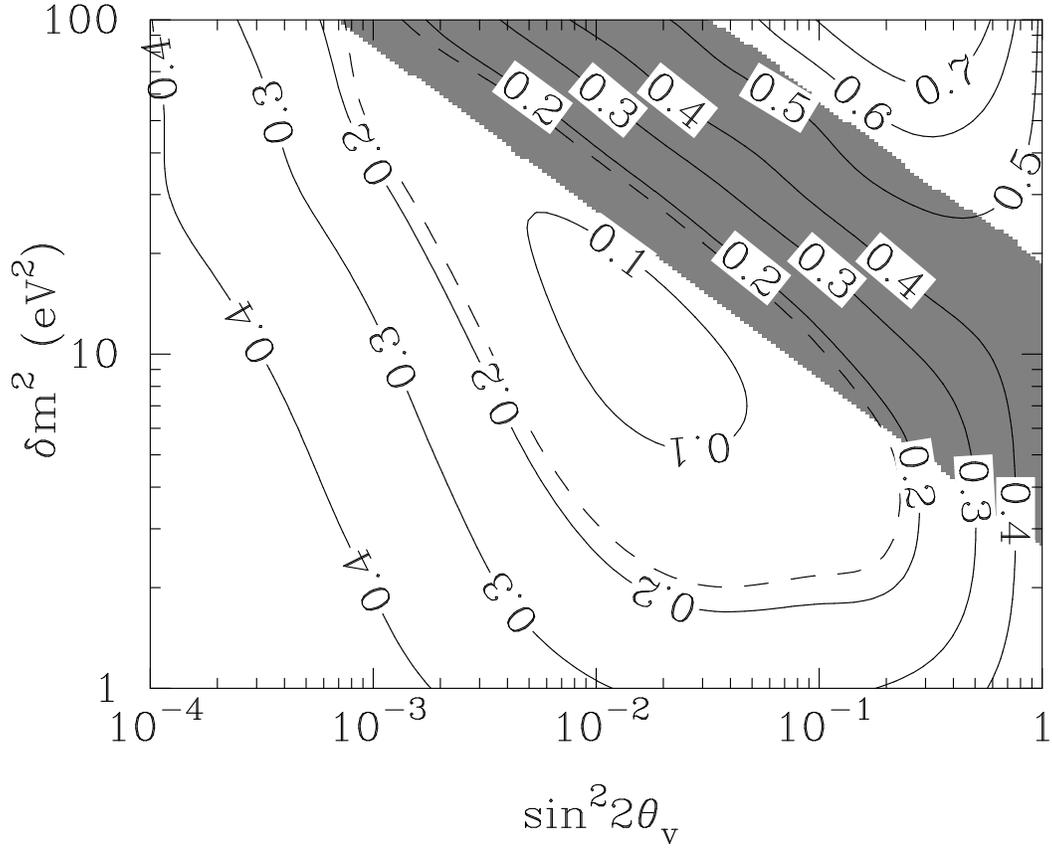}}
\caption{Contour plot of electron fraction as measured at the
point where heavy nuclei begin to form  Neutrino driven wind parameters
employed here are $s/k = 100$, $\tau = 0.3 {\rm s}$.}
\label{fig:efraction2}
\end{figure}

\begin{figure}
\includegraphics[angle=0,width=14cm]{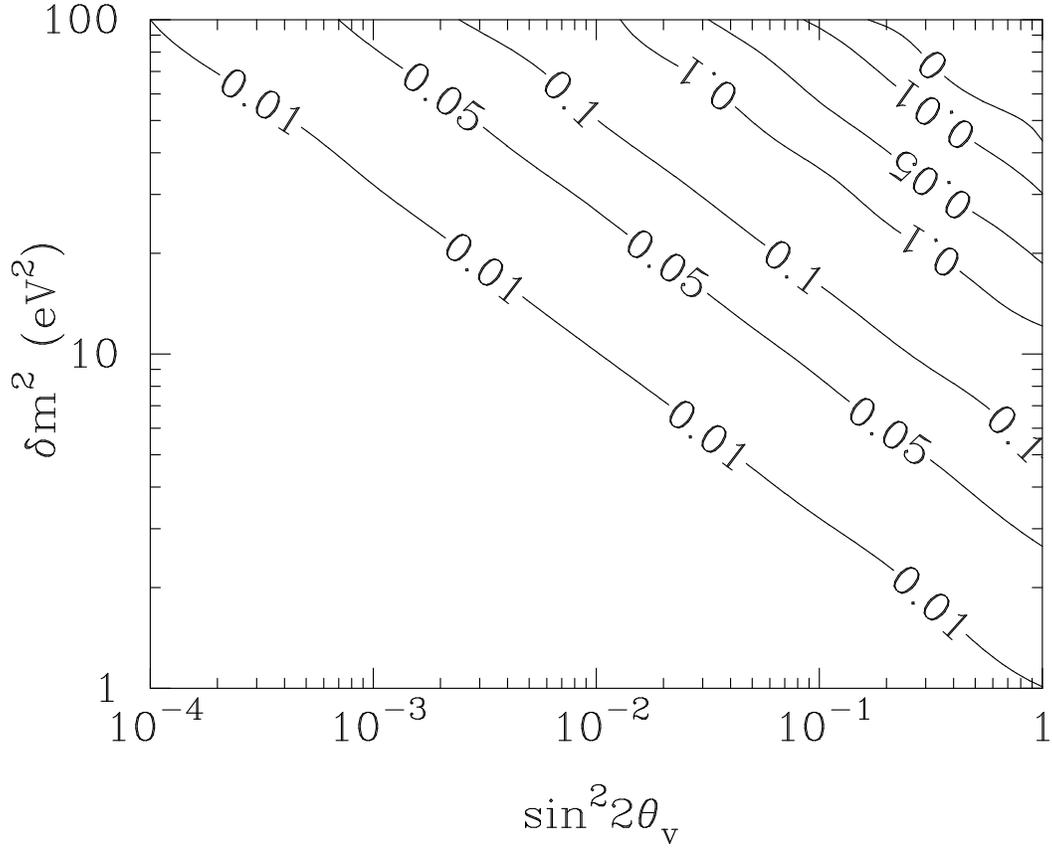}
\caption{Shows the effect of including non radial paths.  The same
wind parameters are used here as in Fig. \ref{fig:efraction2}.  This
shows the difference in $\nu_e$ survival probability at 11 km
with and without nonradial neutrino paths.  For large $\delta
m^2 \sin^2 2 \theta_v$ (the upper right corner), most neutrinos
convert in the radial scenario, so the radial and nonradial results
converge.} 
\label{fig:nonradial}
\end{figure}

\begin{figure}
\includegraphics[angle=-90,width=14cm]{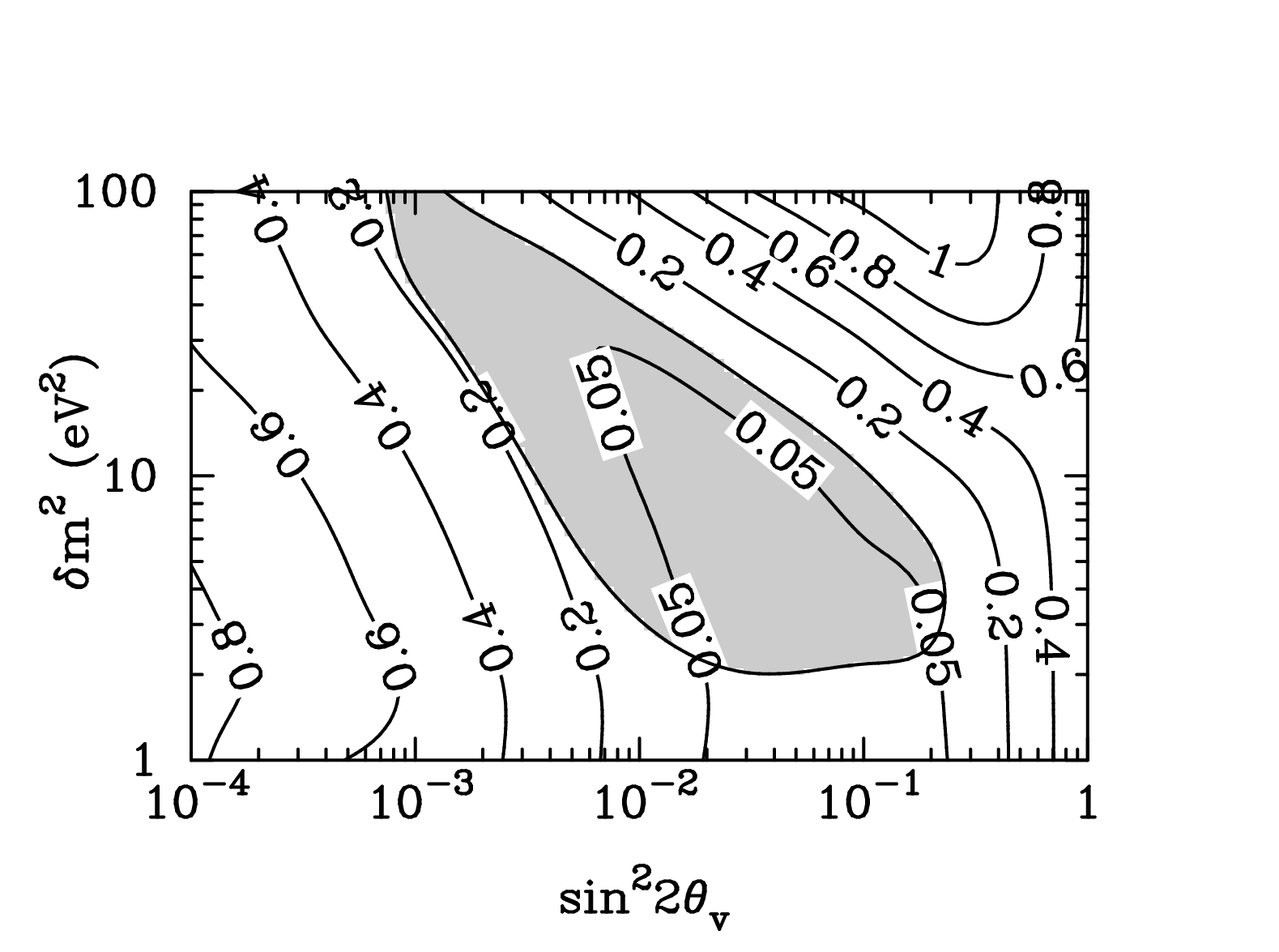}
\caption{
Ratios of events that would be seen in SNO from charged
current electron neutrino break-up of the deuteron with active-sterile
transformation to without active-sterile transformation.}
\label{fig:sno1}
\end{figure}

\begin{figure}
\centerline{\includegraphics[angle=-90,width=14cm]{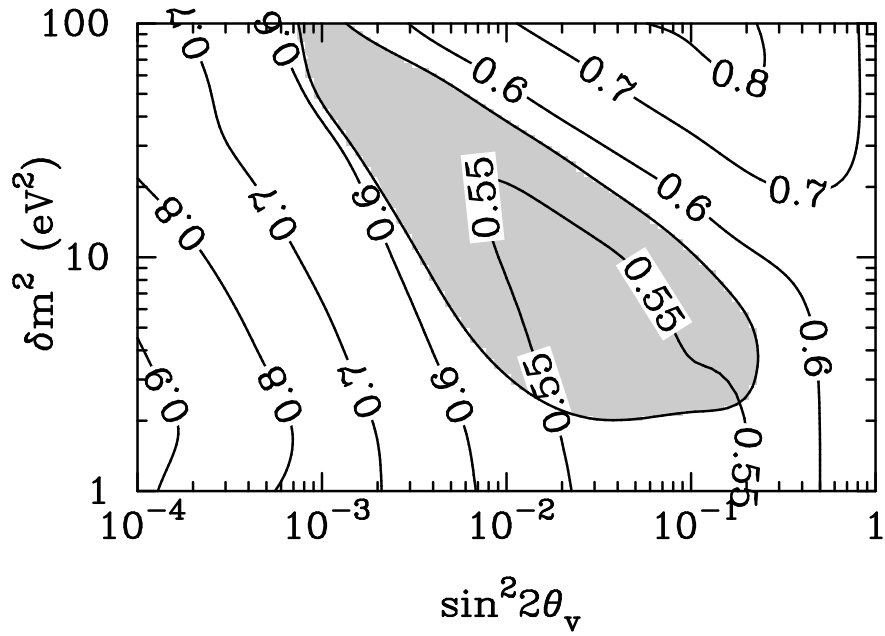}}
\caption{Ratios of events that would be seen in SuperKamiokande from
all types of neutrino scattering on electrons with active-sterile
transformation to without active-sterile transformation.}
\label{fig:elscat1}
\end{figure}

\begin{figure}
\centerline{\includegraphics[angle=-90,width=14cm]{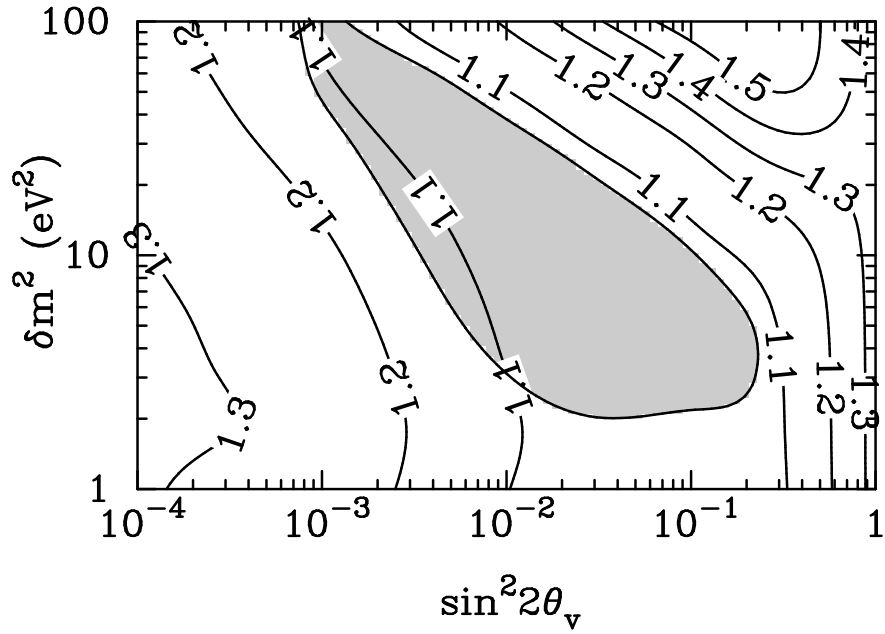}}
\caption{Ratios of events that would be seen in SNO from charged
current electron neutrino break-up of the deuteron with both active-sterile
transformation and the active-active SMA solar solution  to
without any transformation.}
\label{fig:sno2}
\end{figure}

\begin{figure}
\centerline{\includegraphics[angle=-90,width=14cm]{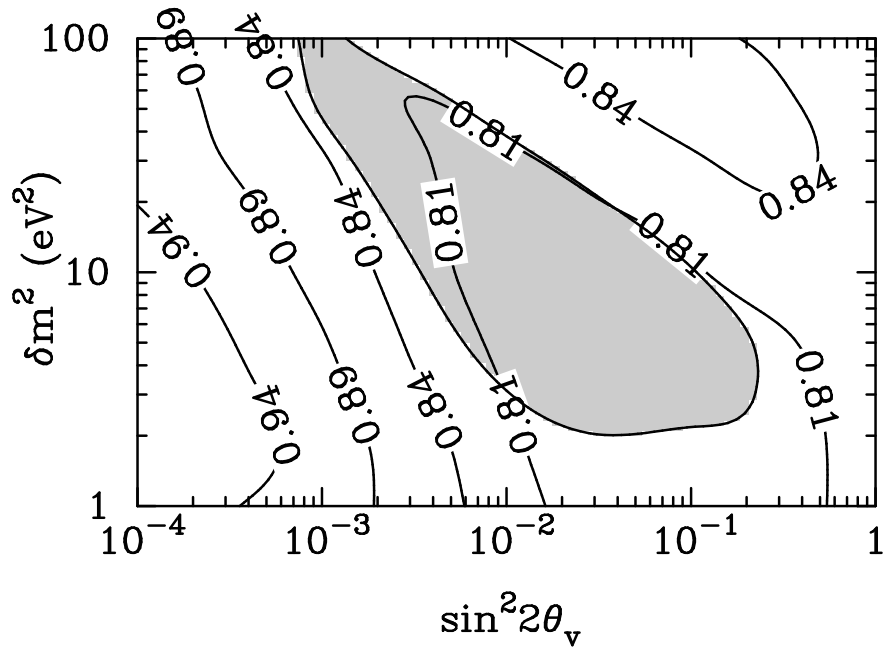}}
\caption{Ratios of events that would be seen in SuperKamiokande from
all types of neutrino scattering on electrons with both active-sterile
transformation and the active-active SMA solar solution to
without any transformation.}
\label{fig:elscat2}
\end{figure}

\newpage

\begin{figure}
\vspace*{-0.5cm}
\includegraphics[width=12cm]{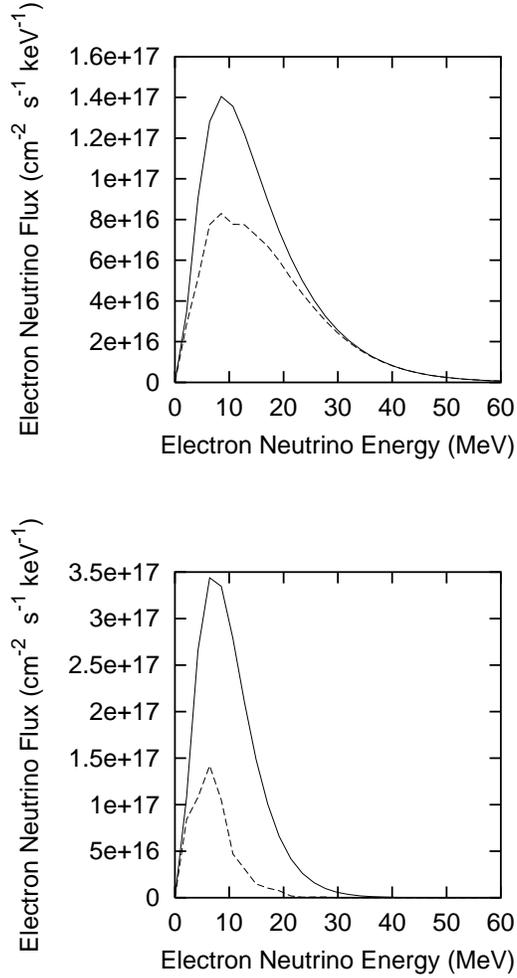}
\vspace*{-1cm}
\caption{The bottom panel shows an example of early (solid line) and late
(dashed line) electron neutrino spectrum for the active-sterile transformation
solution to the r-process. No other transformations with the electron neutrino
are considered.  The top panel shows an example of
the early (solid line) and late (dashed line) electron neutrino spectrum
when both the $\nu_e \leftrightarrow \nu_s$ and the LMA solution to the
solar neutrino problem are considered.  In this panel we assume completely
adiabatic transformation through the $\delta m_{12}$ resonance region.  The
signature of this type of active sterile transformation is a relative
decrease in the number of low energy neutrinos from early to late
times when compared with the number of high energy neutrinos.  }
\label{fig:earlylate}
\end{figure}

\newpage

\begin{figure}
\includegraphics[angle=0,width=14cm]{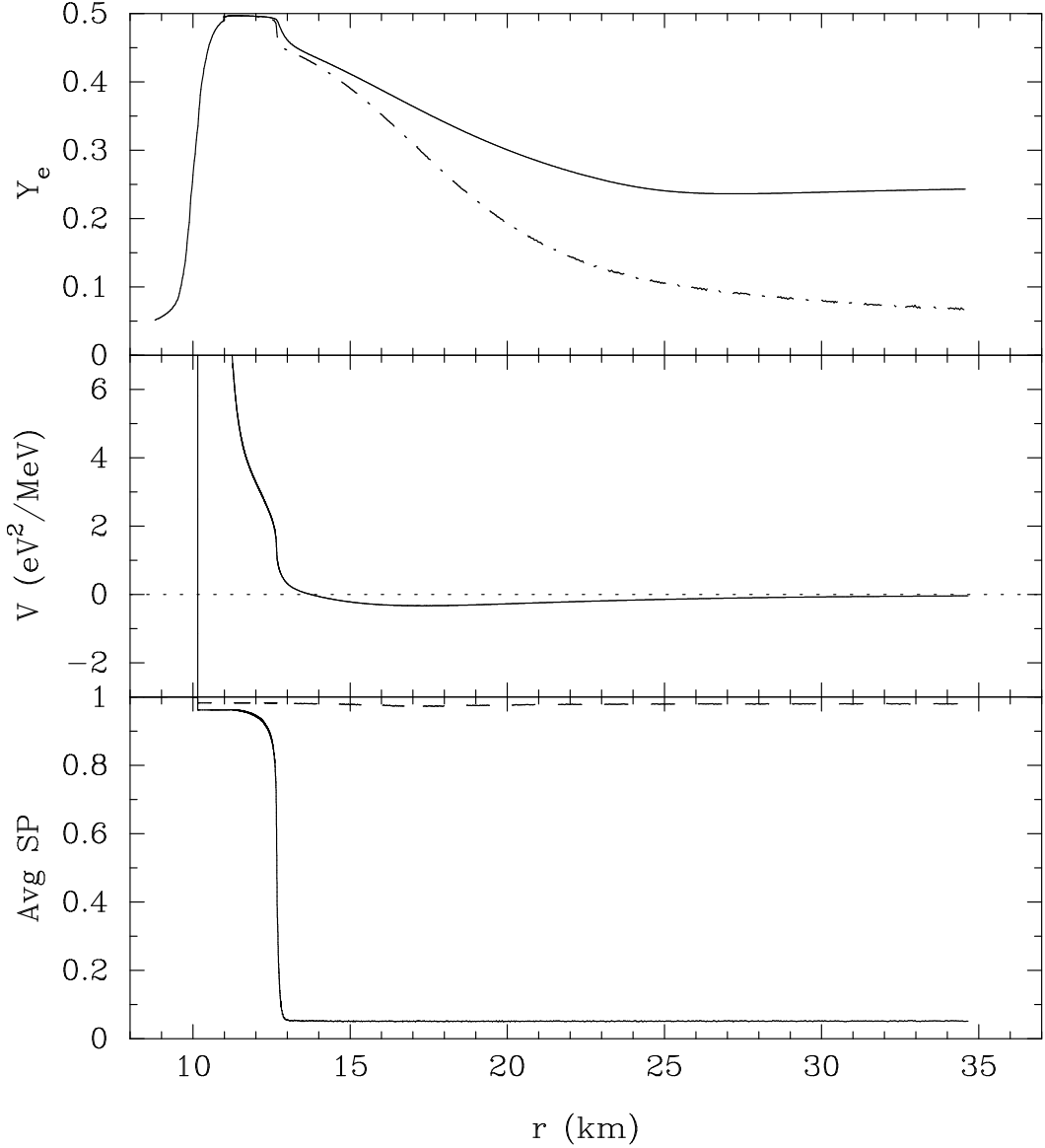}
\caption{ A sample run with the background turned on.  The top
panel shows the actual (solid line) and equilibrium (dashed line)
electron fraction.  The middle panel shows the MSW potential and
the bottom panel shows the average survival probabilities for neutrinos
(solid line) and antineutrinos (dashed line).
The parameters used in this run were  $\delta m^2 = 20 \; {\rm eV}^2$ and
$\sin^ 2 \theta_v = 10^{-2}$. The neutrino luminosities were
$1.08 \times 10^{50}
{\rm erg} {\rm s}^{-1}$ for the neutrinos and
$1.3 \times 10^{50} {\rm erg} \, {\rm s}^{-1}$ for the antineutrinos.
}
\label{fig:back}
\end{figure}

\end{document}